\documentclass[twocolumn,aps,prb,groupedaddress,superscriptaddress,amsmath,floatfix,amssymb,showpacs,noeprint,english]{revtex4-2}
\usepackage{graphicx}
\usepackage{dcolumn}
\usepackage{bm}
\usepackage[T1]{fontenc}
\usepackage{mathrsfs}
\usepackage{amsbsy}
\usepackage{amstext}
\usepackage{amsmath, amsthm, amsfonts, amssymb, bbold}
\usepackage{setspace}
\usepackage{esint}
\usepackage{xcolor,colortbl}
\usepackage{float}
\usepackage[colorlinks,citecolor=blue,urlcolor=blue,linkcolor=blue,hypertexnames=true]{hyperref} 
\usepackage{babel}
\usepackage{booktabs}
\usepackage[section]{placeins}
\usepackage{tikz}

\usepackage{quantikz}

\newcommand{\upa}[0]{\uparrow \,}
\newcommand{\dna}[0]{\downarrow \,}

\begin{document}

\title{Using a Feedback-Based Quantum Algorithm to Analyze the Critical Properties of the ANNNI Model Without Classical Optimization}

\author{G. E. L. Pexe}
\altaffiliation{These authors contributed equally to this work}
\affiliation{Faculty of Sciences, UNESP - S{\~a}o Paulo State University, 17033-360 Bauru-SP, Brazil}

\author{L. A. M. Rattighieri}
\altaffiliation{These authors contributed equally to this work}
\affiliation{Faculty of Sciences, UNESP - S{\~a}o Paulo State University, 17033-360 Bauru-SP, Brazil}

\author{A. L. Malvezzi}
\affiliation{Faculty of Sciences, UNESP - S{\~a}o Paulo State University, 17033-360 Bauru-SP, Brazil}

\author{F. F. Fanchini}
\email{felipe.fanchini@unesp.br}
\affiliation{Faculty of Sciences, UNESP - S{\~a}o Paulo State University, 17033-360 Bauru-SP, Brazil}
\affiliation{QuaTI - Quantum Technology \& Information, 13560-161 São Carlos-SP, Brazil}

\date{\today}

\begin{abstract}

We investigate the critical properties of the Anisotropic Next-Nearest-Neighbor Ising (ANNNI) model using a feedback-based quantum algorithm (FQA). We demonstrate how this algorithm enables the computation of both ground and excited states without relying on classical optimization methods. By exploiting symmetries in the algorithm initialization, we show how targeted initial states can increase convergence and facilitate the study of excited states. Using this approach, we study the quantum phase transitions with the Finite Size Scaling method, analyze correlation functions through spin correlations in the ground state, and examine magnetic structure by calculating structure factors via the Discrete Fourier Transform. Our findings highlight FQA’s potential as a versatile tool for studying not only the ANNNI model but also other quantum systems, providing insights into quantum phase transitions and the magnetic properties of complex spin models.

\end{abstract}

\maketitle

\section{Introduction}

The theoretical and experimental analysis of many-body quantum systems is a widely studied topic \cite{doi:10.1126/science.aag2302, doi:10.1126/science.aaf8834, PhysRevLett.119.073002, PhysRevX.7.041047, PhysRevA.95.053624}. In particular, calculating the ground and excited states of Hamiltonians in these systems is intrinsically linked to elements of interest analyzed through quantum simulation algorithms. Algorithms that enable the acquisition of these states are a central focus of quantum simulation research \cite{larsen2023feedbackbased, Poulin_2009, tubman2018postponing, ge2018faster, Lin_2020, sierant2024manybodylocalizationageclassical}. Despite the inherent complexity of searching for generic ground states, even on quantum computers \cite{kempe2005complexity}, recent advances in technology have ushered in the era of noisy intermediate-scale quantum (NISQ) computers \cite{Preskill_2018}. This progress has driven substantial research into the application of NISQ devices, particularly in quantum simulation \cite{Cerezo_2021, Bharti_2022}, focusing on determining ground states of physical, chemical, and material systems using the Variational Quantum Eigensolver (VQE) Algorithm \cite{Peruzzo_2014, Tilly_2022, kandala2017hardware} and the Quantum Approximate Optimization Algorithm (QAOA) \cite{qaoa}. However, VQE and QAOA face a significant practical challenge, the need for classical optimization of quantum circuit parameters to minimize an objective function, a task that becomes increasingly complex as the search space dimension grows.

In this study, we explore an alternative approach that eliminates the need for optimization, thereby avoiding this complexity. Unlike classical optimization of a parameterized circuit, our proposed strategy utilizes a feedback law principle to sequentially establish quantum circuit parameter values, layer by layer, based on feedback from measurements of qubits in the preceding layer. This feedback principle is based on Lyapunov quantum control theory \cite{Cong_2013} and has been developed to ensure that the objective function value monotonically decreases with circuit depth. This approach represents an innovative application of Feedback-based Quantum Algorithm (FQA), extending principles from the Feedback-based Quantum Optimization Algorithm (FALQON) \cite{larsen2023feedbackbased, 1Magann_2022, 2Magann_2022}, recently developed for combinatorial optimization. 

This work focuses on advancing FQA to efficiently reach excited states by exploiting model symmetries. We show that the use of symmetries reduces the search space by confining it to specific sectors defined by quantum numbers associated with symmetry operators \cite{malvezzi2003introduction, Jung_2020}. This reduction not only guides the selection of initial states but also ensures that these states remain within the symmetry sector of the desired target state. In fact, we show that a proper alignment with the symmetry sector minimizes overlap with other subspaces, thereby enhancing the algorithm's convergence towards the target state. Our results establish a framework that can be generalized to other systems, enabling a more efficient preparation of excited states in FQA and similar algorithms. Moreover, this approach provides a pathway for extending the applicability of FQA beyond the ANNNI model, contributing to the study of phase diagrams, correlation functions, and the structural characteristics of complex quantum systems.

With the tools developed to efficiently calculate both ground and excited states using FQA, a detailed exploration of the ANNNI model becomes feasible. Investigating the phase diagram provides insights into the critical properties and magnetic behavior of matter at microscopic levels. By employing Finite Size Scaling (FSS), which requires access to excited states, quantum phase transitions can be characterized. Furthermore, structural factors can be examined through the discrete Fourier transform, enabling the study of spin oscillation patterns at various wavelengths and the spatial organization of magnetic moments along the chain. These analyses offer a comprehensive understanding of the ANNNI model and its intricate quantum phenomena.

To develop these studies, this paper is organized as follows: Section \ref{annni} provides a review of the ANNNI model Hamiltonian, discussing its construction and phase diagram. Section \ref{FALQON} offers an in-depth review of FQAs. The results of our quantum simulations, along with a detailed analysis of the ANNNI model, are presented in Section \ref{results}. Conclusions and future perspectives are discussed in Section \ref{concl}.

\section{ANNNI Model}
\label{annni}
\subsection{Hamiltonian}

The ANNNI model is a fundamental theoretical model in statistical physics, describing the behavior of spin-$1/2$ particles in a one-dimensional magnetic system. This model is particularly interesting due to its anisotropy in interactions between nearest and next-nearest neighbor spins \cite{PhysRevLett.44.1502, SELKE1988213, Sa_Barreto_2002, Barone_2024}. Here, we are interested in the stability of phases present in the model and the phase transitions driven by quantum fluctuations at zero temperature. Quantum fluctuations can be further influenced by the presence of a transverse field. Therefore, we consider the properties of the ground state of the ANNNI model in a transverse field at zero temperature. Frustration is expected to arise due to the presence of next-nearest neighbor interactions, leading to interesting quantum phases. The model can be described by the following Hamiltonian:

\begin{equation}
\label{H}
H_{p} = -J\sum_{j = 1}^{L}\left ( \sigma _{j}^{z}\sigma _{j+1}^{z}-\kappa\sigma _{j}^{z}\sigma _{j+2}^{z}+g\sigma _{j}^{x} \right ),
\end{equation}

\noindent
where $\sigma _{j}^{a}$, with $a = x, y, z$, are the Pauli matrices acting on site $j$ of a one-dimensional lattice (chain) with $L$ sites. For a chain with $L$ sites, the Hamiltonian $H$ acts in a Hilbert space of dimension $2^L$, which is the tensor product of $L$ two-dimensional spaces. The coupling constant $J (>0)$ represents the ferromagnetic interaction strength between nearest neighbors, defining the energy scale (we set $J = 1$), while $\kappa (>0)$ and $g (>0)$ are dimensionless coupling constants associated with next-nearest neighbor interactions and the transverse magnetic field, respectively.

\subsection{Phase Diagram} 

\begin{figure}[htpb]
    \centering
    \includegraphics[width=0.9\linewidth]{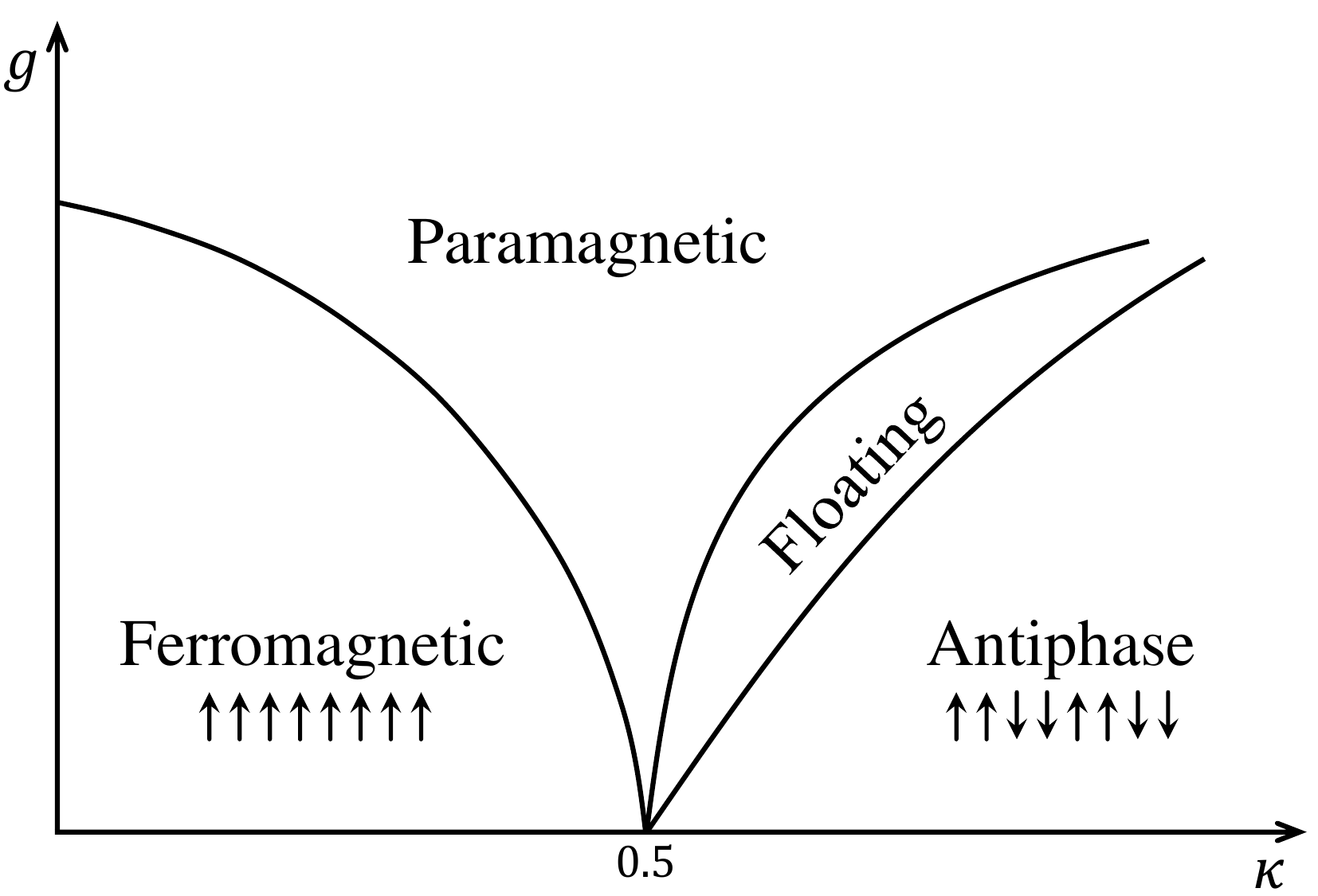}
    \caption{Schematic phase diagram of the transverse ANNNI model. \cite{Suzuki, PhysRevE.75.021105, mattis1985theory}}
    \label{fig:phase_diagram}
\end{figure}

The phase diagram of the ground state of the ANNNI model reveals four distinct phases: ferromagnetic, antiferromagnetic, paramagnetic, and incommensurate floating, each separated by three quantum phase transitions, as shown in Fig. \ref{fig:phase_diagram}.

Understanding the phases of the ANNNI model is relevant for analyzing how the microscopic two-body interactions manifest themselves on a many-body scale \cite{ferreiramartins2023detecting, cea2024exploring, SELKE1988213}. In the ANNNI model, spins of nearest neighbors interact ferromagnetically, favoring alignment in the same direction, while spins of next-nearest neighbors interact antiferromagnetically, causing them to align in opposite directions. When the interaction between nearest neighbors dominates, the system is in the ferromagnetic phase. When the interaction between next-nearest neighbors becomes more significant, the system exhibits a compromising solution, characterized by  the antiphase \cite{yang2023pattern}. The inclusion of a transverse field in the ANNNI model tends to induce disorder, resulting in the paramagnetic phase, where the spins do not exhibit a defined alignment with respect to the $z$ direction.

\section{FQA: Feedback Quantum Algorithm}
\label{FALQON}

\begin{figure*}[htpb]
    \centering
    \includegraphics[width=1\linewidth]{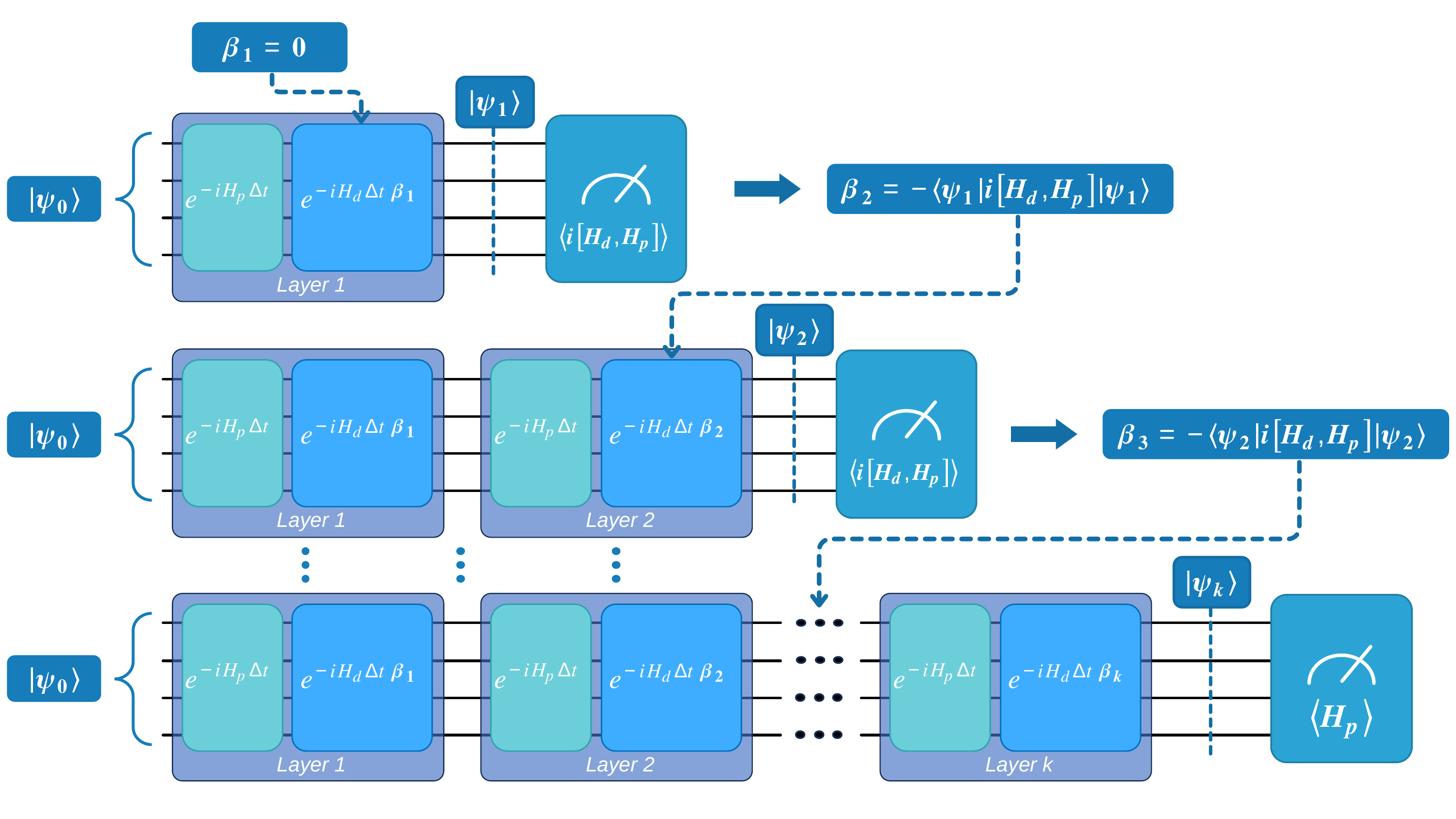}
    \caption{Illustrative diagram of the FQA quantum algorithm. The process starts with an initial state $\ket{\psi_0}$. At each layer $k$, the unitary operators $e^{-iH_p \Delta t}$ and $e^{-iH_d \Delta t \beta_k}$ are sequentially applied to the current state, adaptively adjusting the parameter $\beta_k$. This process is iteratively repeated to approximate the target state, representing the evolution of the state $\ket{\psi}$ across the layers until reaching the desired solution.}
    \label{fig:enter-label}
\end{figure*}

In this section, we introduce the concept of Feedback-based Quantum Algorithms (FQAs) and their relationship with the continuous Lyapunov quantum control framework. This approach is aligned with the content presented by \cite{1Magann_2022}, who introduced FALQON as an FQA for solving combinatorial optimization problems. Additionally, we further develop FQA to find ground and excited states of the ANNNI model. 
As proposed by Magann et al. in their manuscript \cite{1Magann_2022, 2Magann_2022}, we begin with a quantum system whose dynamics are governed by 

\begin{equation}
    i \frac{d}{dt} \ket{\psi(t)} = (H_p + H_d \beta(t)) \ket{\psi(t)}.
\end{equation}

The general solution of this equation is expressed as:

\begin{equation}
|\Psi(t)\rangle = \mathcal{T} e^{-i\int_{0}^{t}(H_{p}+H_{d}\beta(t'))dt' }|\Psi(0)\rangle,
\label{se6}
\end{equation}

\noindent
where $\mathcal{T}$ denotes the time ordering operator, $H_d$ is referred to as the control Hamiltonian, and $\beta(t)$ is a time-dependent control function. The goal of the numerical method is to infer a dynamic by adjusting $\beta(t)$ such that the resulting quantum state tends towards the ground state of $H_p$.

Before outlining the strategy to compute the appropriate function for $\beta(t)$, we will first demonstrate how to implement the aforementioned dynamics in a quantum circuit. For this purpose, we discretize the evolution into two steps. First, time is divided into a sequence of $l$ steps, and the Hamiltonian is approximated as time-independent in each step:

\begin{equation}
|\Psi(t+ \Delta t)\rangle = e^{-i(H_{p}+\beta(t)H_{d})\Delta t} |\Psi(t)\rangle.
\end{equation}

The second step involves approximating the evolution in each time step using trotterization.

This sequence of approximations aims to decompose the complex unitary operation into a product of simpler exponentials:

\begin{widetext}
\begin{equation}
    |\Psi_{l}\rangle = e^{-i\beta _{l}H_{d}\Delta t}e^{-iH_{p}\Delta t}\cdots  e^{-i\beta _{1}H_{d}\Delta t}e^{-iH_{p}\Delta t}|\Psi_{0}\rangle =U_{d}(\beta _{l})U_{p}\cdots U_{d}(\beta _{1})U_{p}|\Psi_{0}\rangle, 
    \label{se9}
\end{equation}
\end{widetext}

\noindent
representing the discretization of Eq. (\ref{se6}) into a form suitable for study on a quantum computer. Here, $\beta_{k}\equiv \beta (k\Delta t)$ with $k = 1, 2, \dots, l$, $|\Psi_{k}\rangle \equiv |\Psi(k\Delta t)\rangle$,  
$U_{p}\equiv e^{-iH_{p}\Delta t}$, and $U_{d}(\beta_{k})\equiv e^{-i\beta_{k}H_{d}\Delta t}$. This provides us with a straightforward method to implement the desired dynamics on a quantum computer.

Given this approach, we will now consider an appropriate form for $\beta(t)$, a function that drives the initial state to the ground state of $H_p$. We seek $\beta(t)$ in a way that 

\begin{equation}
\frac{d}{dt} \langle \psi(t) | H_p | \psi(t) \rangle (t) \leq 0, \quad \forall t \geq 0.
\end{equation}

Indeed, there are various functions for $\beta(t)$ that can guide the quantum state to the ground state, and one possible choice is given by $\beta(t) = -i\langle\Psi(t)|[H_{d},H_{p}]|\Psi(t)\rangle$ \cite{1Magann_2022}. Given this, we can now describe the mechanics of implementing FQA as illustrated in Fig. \ref{fig:enter-label}. The first step is to select an initial state $|\Psi_{0}\rangle$ and a time step $\Delta t$, and start the algorithm with a value for $\beta_{1}$. FQAs then use the feedback law:

\begin{equation}
\beta_{k} = -A_{k-1}, \label{se10}
\end{equation}

where $A_{k} = i \langle\Psi_{k}|[H_{d},H_{p}]|\Psi_{k}\rangle$,
to determine the values of the quantum circuit parameters $\beta_{k}$ at steps $k = 2, \dots, l$. The term in Eq. (\ref{se10}) is termed as the feedback law because at each step, $A_{k-1}$ is ``fed back'' to establish the subsequent value of $\beta_{k}$. For sufficiently small $\Delta t$, this procedure ensures that the cost function $J_{k} = \langle\Psi_{k}|H_{p}|\Psi_{k}\rangle$ monotonically decreases with respect to layer $k$, i.e., $J_{1}\geq J_{2} \geq \cdots \geq J_{l} $, following a discretized version of Eq. (\ref{se3}). It is important to note that a positive limit for $\Delta t$ ensuring the maintenance of this property can be obtained based on the steps described in Appendix A of \cite{deltat}, and is expressed in terms of the spectral norms of $H_p$ and $H_d$ as

\begin{equation}
\Delta t \leq \frac{1}{4\left \| H_p \right \| \left \| H_d \right \|^{2}} . \label{se12}
\end{equation}

In each step $k$ of an FQA implementation, the quantum state $|\Psi_{k}\rangle$ is prepared by applying the $k$-layer quantum circuit $U_{k}(\beta_{k})U_{p}\cdots U_{d}(\beta_{1})U_{p}$ to the fixed initial state $|\Psi_{0}\rangle$. At the end of this circuit, the value of $A_{k}$ can be estimated, for example, through repeated measurements of the observable $i[H_{d}, H_{p}]$, decomposed into a linear combination of Pauli operators, in the state $|\Psi_{k}\rangle$.

\section{Results}
\label{results}
In this section, we show how a quantum computer can be used to study complex magnetic systems using FQA. This section is divided into four subsections. First, we demonstrate how to use FQA to calculate the ground and excited states. Following this, we illustrate how this tool can be employed to study quantum phase transitions, spin correlation functions, and the magnetic structure factor.

\subsection{Calculating Ground and Excited State Energies Using FQA}

To use FQA as a tool to study magnetic systems, we consider the states $\ket{0} = \ket{\uparrow}$ and $\ket{1} = \ket{\downarrow}$, where $\ket{\uparrow}$ and $\ket{\downarrow}$ are the eigenstates of the spin operator $\sigma^z$, with eigenvalues $+1$ and $-1$, respectively. Based on these states, we define the Hamiltonian $H_d$ as:

\begin{equation}
H_d = \sum^{L}_{j=1} \sigma_j^x.
\end{equation}

\begin{figure}[h!]
\centering
\includegraphics[width=1\linewidth]{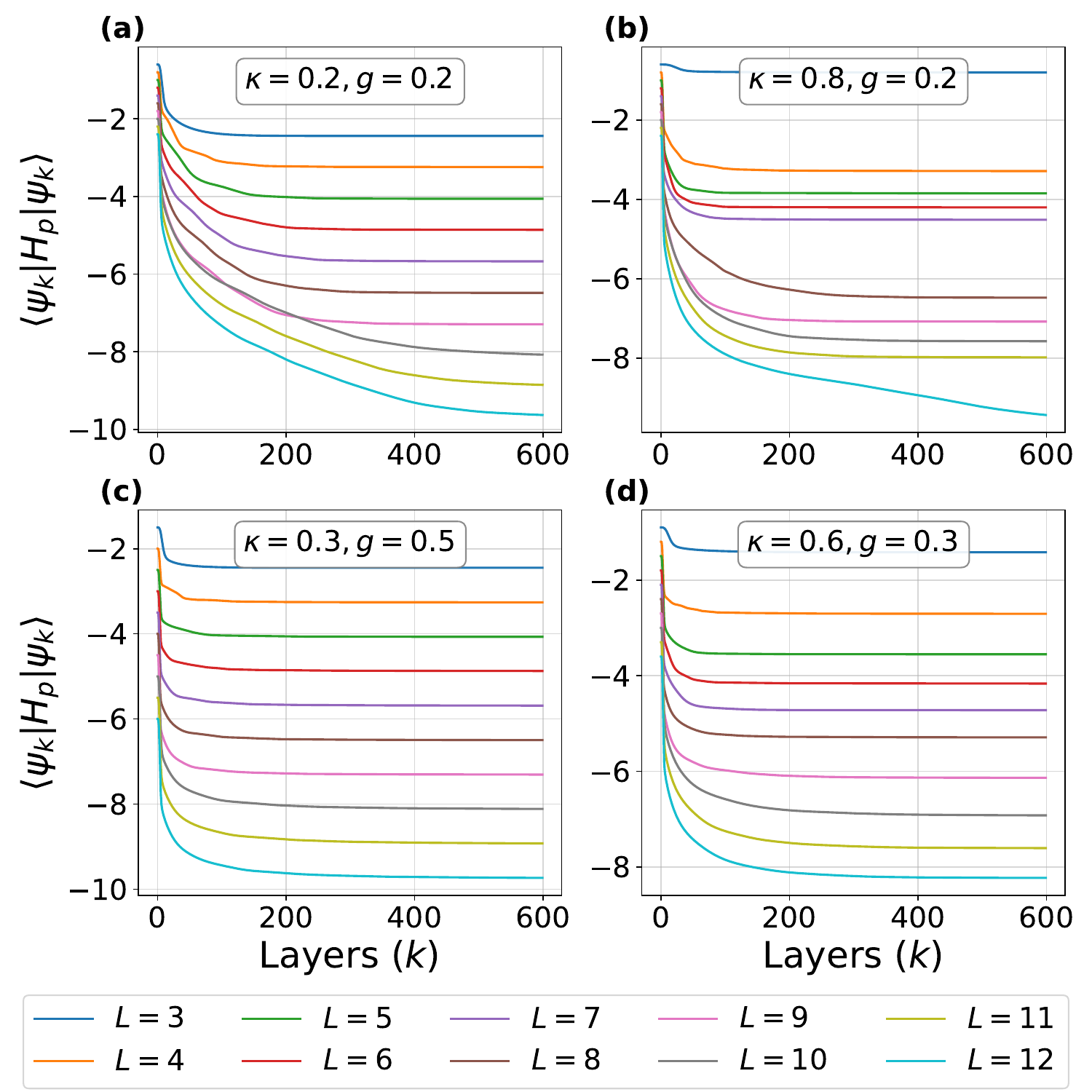}  \caption{Numerical simulations of FQA applied to the ANNNI model for chains of 3 to 12 sites with different values of $\kappa$ and $g$. Simulations use $\Delta t = 0.02$. In the panels, the curves show the convergence of the cost function, $J = \langle\Psi_{k}|H_p|\Psi_{k}\rangle$, as a function of layer $k$, when qubits are initialized in the ground state of $H_{d}$.}
\label{fig:universe}
\end{figure}

This is the standard way the mixer is defined in the literature. Although it is not strictly necessary to define it this way, we adhere to this convention in our work. Figure \ref{fig:universe} shows the numerical simulations of FQA applied to the ANNNI model for chains of 3 to 12 sites, with different values of $\kappa$ and $g$. The simulations use $\Delta t = 0.02$. Each panel illustrates the convergence of the cost function, $J = \langle \Psi_k | H_p | \Psi_k \rangle$, as a function of the number of layers, for different chain lengths. The expected value of the objective function monotonically decreases with the number of layers, demonstrating the convergence of FQA to the ground state, even for values close to the transition lines where quantum fluctuations are more intense.

It is worth noting that the value of $\Delta t$ is directly related to the chain size $L$, and an appropriate choice results in satisfactory performance. Good performance can be achieved in practice for $\Delta t$ values significantly exceeding the limit presented in Eq. (\ref{se12}). However, when $\Delta t$ is too large, we observe that the algorithm does not converge. In this regime, the parameters $\beta_k$ oscillate drastically, and the objective function does not decrease monotonically (for more details, see \cite{larsen2023feedbackbased, 2Magann_2022}). In practice, FQA does not require evaluations of the objective function at every step; it only needs the estimation of $A_k$ to define each $\beta_k$ value. Thus, signs of oscillatory behavior in $\beta_k$ (or equivalently, in the record of $A_k$ measurements) may indicate that $\Delta t$ is too large and the algorithm is not converging.

As mentioned earlier, our analysis demonstrates that, beyond the ground state, FQA can also be utilized to calculate the energy of excited states, provided the algorithm incorporates the symmetry of the initial state. Notably, when the problem Hamiltonian ($H_p$) and the driver Hamiltonian ($H_d$) share the same symmetries, as is the case in this study, the initial state exhibits a remarkable property. If it is chosen within a specific sector with defined quantum numbers, it remains confined to that sector throughout the execution of the FQA. This behavior arises because the time evolution operators associated with $H_p$ and $H_d$ preserve the state within the same symmetry sector (see Appendix \ref{ap:simetrias} for details). Consequently, the algorithm ensures convergence to the lowest-energy eigenvalue of $H_p$ within the specified sector. This approach is particularly effective for calculating excited states of the Hamiltonian, as it allows targeting distinct symmetry sectors without interference from other subspaces.

In this way, the convergence of FQA to the ground state depends on the initial state belonging to the symmetry sector that contains the ground state. This behavior becomes particularly advantageous for calculating excited states, as the algorithm can target specific sectors to isolate these energies. The process involves preparing initial states in different sectors and monitoring the energy to which FQA converges. By comparing the results, the ground state and successive excited states can be systematically identified. Furthermore, insights into the symmetries of the problem can be obtained from analyzing smaller chains, where computational resources are less demanding. These insights can then be used to guide calculations in larger systems, allowing for the determination of excited energies and states more efficiently (for more details and a concrete example, see Appendix \ref{ap:convergence}, where we demonstrate how various excited states can be reached depending on the symmetry of the initial state). Finally, while this strategy is demonstrated here for the ANNNI model, it can be used to other quantum systems, providing a practical framework for studying quantum phase transitions and magnetic properties in a variety of models. 

\begin{figure}[!h]
    \centering
    \includegraphics[width=1\linewidth]{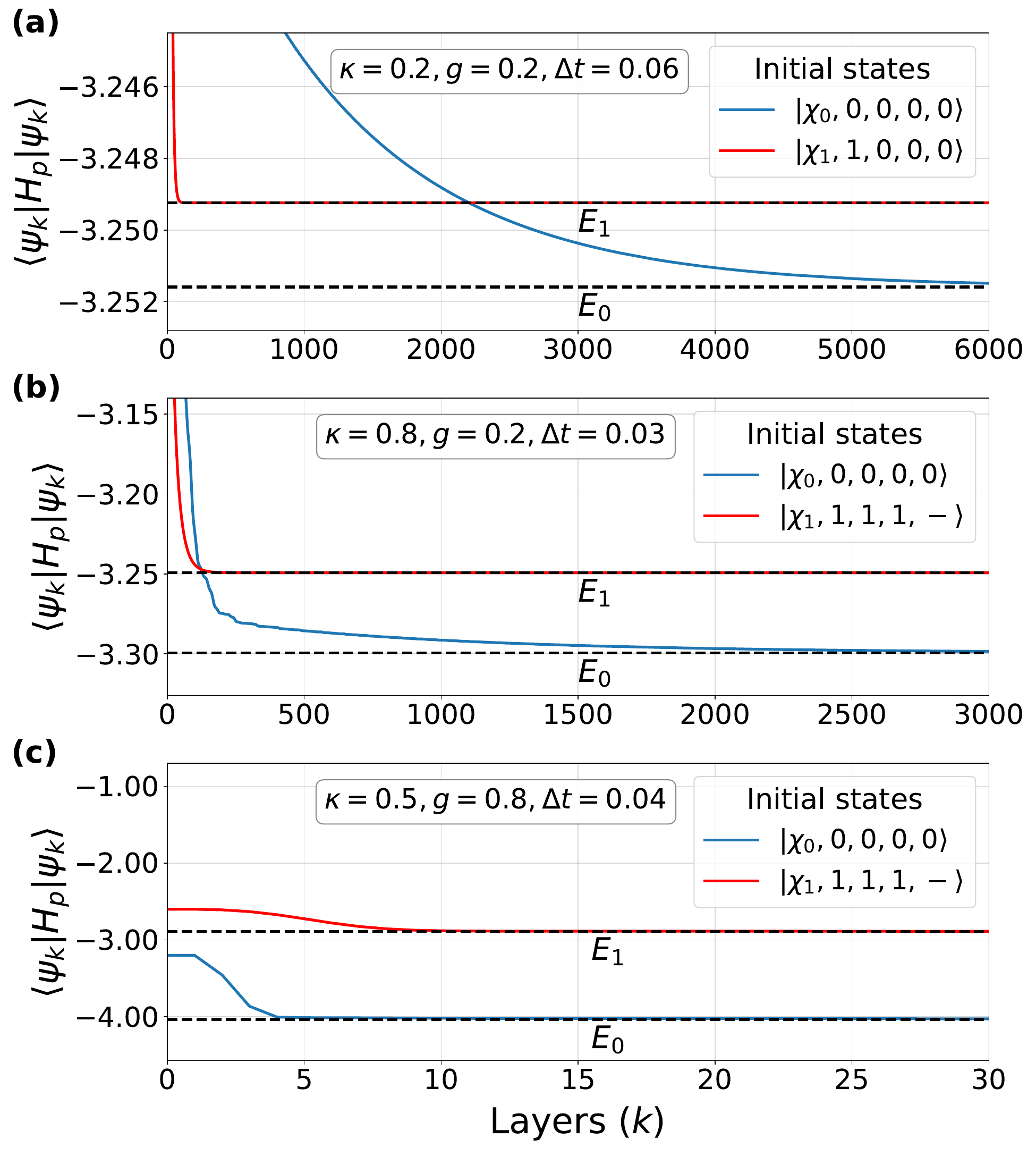}
    \caption{Numerical simulations of FQA applied to the ANNNI model for a chain with 4 sites, for parameters $\kappa = 0.2, g = 0.2$ (a), $\kappa = 0.8, g = 0.2$ (b), and $\kappa = 0.5$ and $g = 0.8$ (c). In the figures, dashed lines indicate the energy levels of the ANNNI Hamiltonian, while curves represent the convergence of the cost function $J = \langle\Psi_{k}|H_p|\Psi_{k}\rangle$ as a function of layer $k$, when qubits are initialized in different eigenstates of $H_{x}$.}
    \label{fig:grafico_convergencia}
\end{figure}

To illustrate the impact of state initialization in different symmetry sectors on the convergence of FQA,  we applied the algorithm to the ANNNI model with a 4-site chain. Figure \ref{fig:grafico_convergencia} presents the convergence results for parameters $\kappa = 0.2, g = 0.2$ [Fig. \ref{fig:grafico_convergencia}(a)], $\kappa = 0.8, g = 0.2$ [Fig. \ref{fig:grafico_convergencia}(b)], and $\kappa = 0.5, g = 0.8$ [Fig. \ref{fig:grafico_convergencia}(c)], corresponding to different phases of the model. 
The notation for the initial states used is explained in detail in Appendix \ref{ap:simetrias}, but it is worth emphasizing that $\chi_i$ specifically refers to the $i$-th lowest energy eigenstate of the Hamiltonian $H_d$.
In Figure, we use different values of $\Delta t$ because the algorithm's convergence varies with the step size, and smaller values of $\Delta t$ require a larger number of layers in the circuit.  Each parameter setting has an ideal $\Delta t$ value, which should be large enough to minimize circuit depth, but not so large that it causes a loss of monotonicity in the energy curve or drastic variations in $\beta_k$ at each layer (for more details, see \cite{larsen2023feedbackbased}). Although it is possible to choose a very small $\Delta t$ that is consistent across all configurations, this approach increases circuit depth and results in additional computational costs. In Figs. \ref{fig:grafico_convergencia}(a)-(c), the state $\ket{\chi_0, 0, 0, 0, 0}$ reaches the ground state of the Hamiltonian, while the state $\ket{\chi_1, 1, 0, 0, 0}$ reaches the first excited state in (a), and $\ket{\chi_1, 1, 1, 1, -}$ in (b), (c). 

\subsection{Quantum Phase Transition}

\begin{figure}[!h]
\centering
\includegraphics[width=1\linewidth]{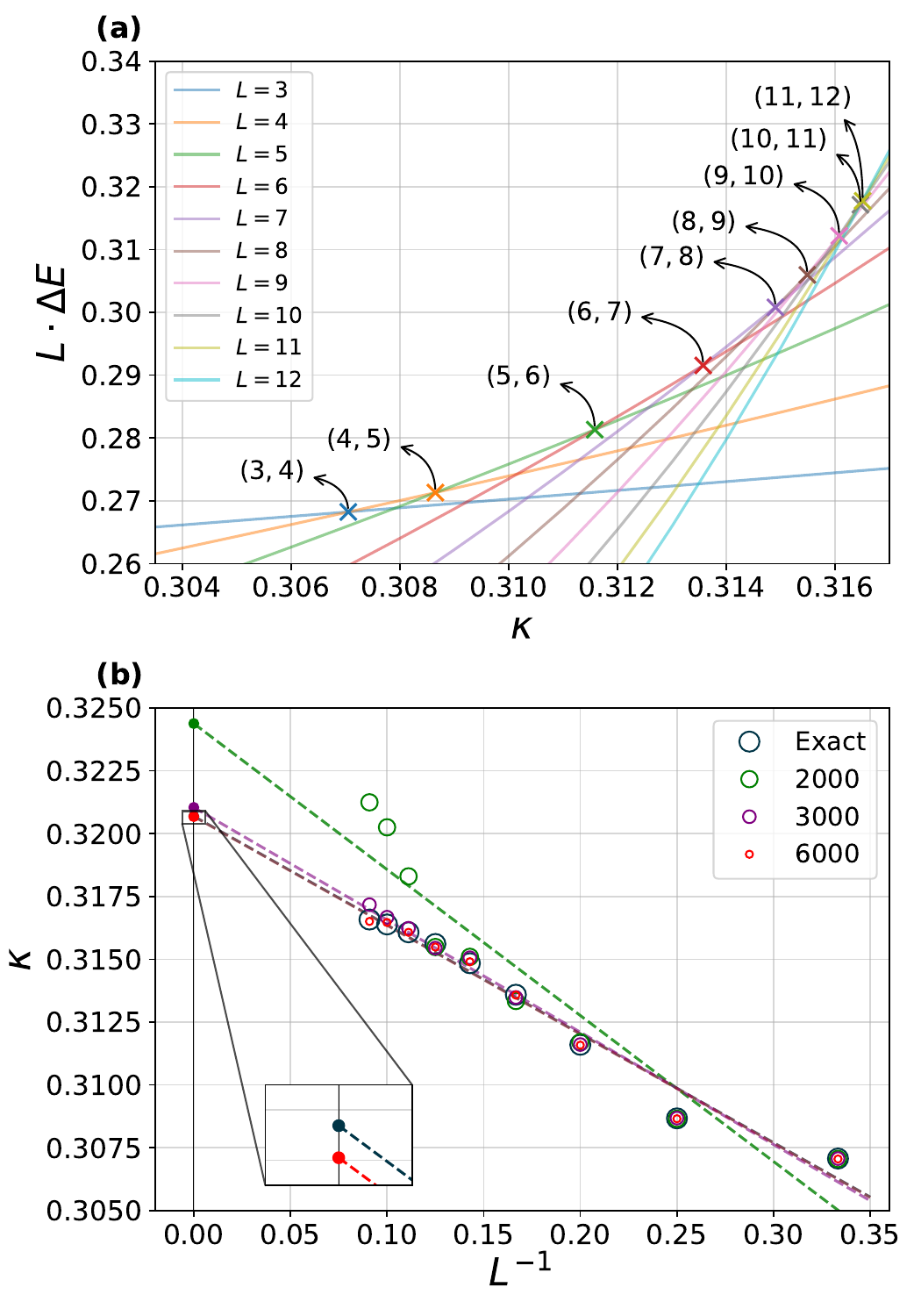}  
\caption{Application of the Finite Size Scaling method using FQA. In (a), the energy gap $\Delta E$ is presented as a function of $\kappa$ for various finite lattices near the critical region, with $g = 0.40$, 6000 layers of FQA, where $\Delta t = 0.03$ was used for chains with fewer than 9 sites and $\Delta t = 0.02$ for larger chains. In (b), the Finite-Size dependence of the parameter $\kappa$ as a function of $\frac{1}{L}$ is shown, as obtained in (a), for different layers of FQA, compared with the exact value. Each point represents the intersection between two adjacent chain sizes, $L$ and $L+1$, where a linear extrapolation to $\frac{1}{L} = 0$ is performed.
}
\label{fig:fss}
\end{figure} 

Since we have presented a strategy to calculate the energy of the ground and excited states of the ANNNI model, we can now explore various properties of the model. We begin by examining the quantum phase transitions. For this purpose, we used the FSS method to detect the phase transition. The FSS method is a technique used to estimate the critical points of quantum phase transitions by analyzing how the energy gap between ground and excited states scales with the size of the system. By calculating this energy gap for different chain sizes and extrapolating the results to the infinite system limit, FSS allows for accurate identification of phase transition points (for more details, see Appendix \ref{ap:FSS}).

To implement this with FQA, we calculated the energy gap between the ground state and the first excited state by varying the chain sizes. {However, applying FSS with the FQA algorithm presents challenges, as the energy difference between the ground state and the first excited state for different chain sizes is often very small, requiring highly precise energy calculations. This implies the need for a large number of layers in FQA and a significant number of measurements. To mitigate these challenges, we can adopt initial conditions that are more favorable to the regime under study, which significantly reduces the algorithm's convergence time. As illustrated in Fig. \ref{fig:graf_convergencia_estados}, we used the initial conditions $(|0\rangle^{\otimes L} + |1\rangle^{\otimes L})/\sqrt{2}$ to calculate the ground state and $(|0\rangle^{\otimes L} - |1\rangle^{\otimes L})/\sqrt{2}$ to calculate the first excited state, both of which respect the symmetry of the ground and excited states of the model. These states perform better in the ferromagnetic phase, as they correspond to the ground states of the model when $\kappa = 0$ and $g = 0$.}

For fixed values of $g$, the parameter $\kappa$ was varied, presenting the energy gap $\Delta E$ as a function of $\kappa$. From the intersection between two adjacent chain sizes, $L$ and $L+1$, we performed a linear extrapolation to $\frac{1}{L} \rightarrow 0$. Figure \ref{fig:fss}(a) depicts the relationship between the energy gap of the first excited state with respect to the ground state ($\Delta E \cdot L$) and the next-nearest-neighbor parameter ($\kappa$) for different chain sizes, with the external field parameter $g$ fixed at $0.4$. 

The behavior of energy gap crossings as a function of the parameter $\kappa$ provides crucial insights into the phase transition of the system under study. FSS method is used to extrapolate these crossings, allowing estimation of the critical point of the phase transition. As $\frac{1}{L}$ tends to zero, the crossings converge to the critical point, marking the phase transition, as illustrated in Fig. \ref{fig:fss}(b), which shows the relationship between the parameter $\kappa$ and $\frac{1}{L}$. The data are plotted for different numbers of FQA layers: 2000, 3000, and 6000 layers, compared against the exact value. Extrapolation to $\frac{1}{L} \rightarrow 0$ is crucial for determining the critical point of the phase transition. The convergence point of the crossings from different datasets provides a reliable estimate of the system's critical point.

\begin{figure}
\centering
\includegraphics[width=1\linewidth]{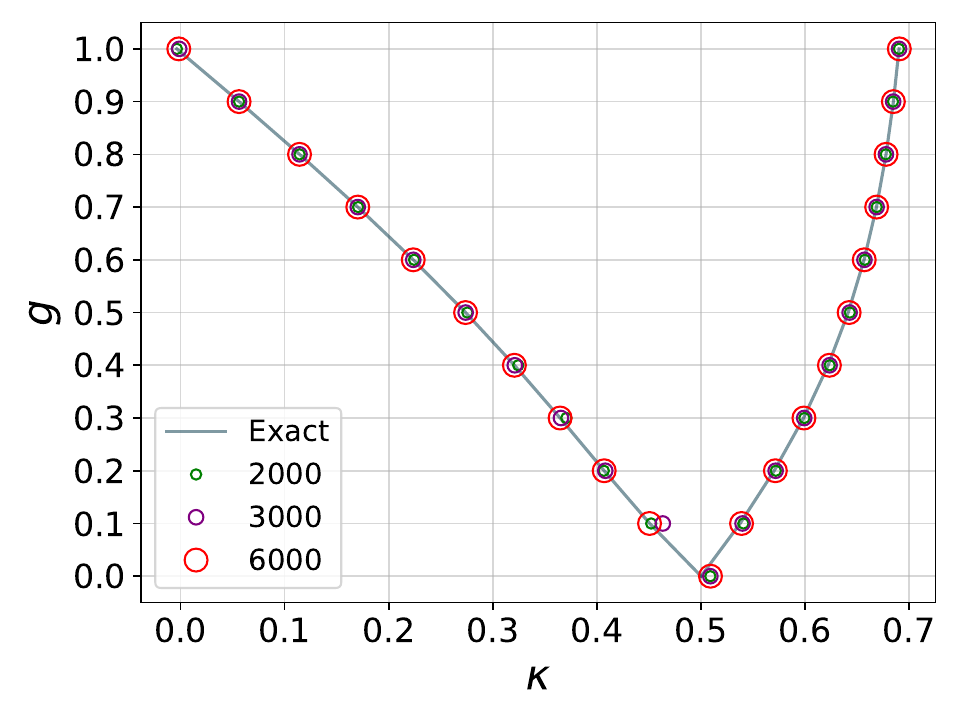}  
\caption{Global extrapolated phase diagram obtained through FQA. For $\kappa < 0.5$, extrapolations are performed as illustrated in Figure \ref{fig:fss}, keeping the FQA parameters consistent with the previous figure. For $\kappa > 0.5$, although a different method is used \cite{Sa_Barreto_2002}, the value of $\Delta t$ has been kept constant at $0.03$.
}
\label{fig:diagFALQON}
\end{figure}

The combined analysis of Fig. \ref{fig:fss}(a) and the extrapolation in Fig. \ref{fig:fss}(b) correctly predicts the phase transition between the ferromagnetic and paramagnetic phases, which occurs for $\kappa < 0.5$ in the phase diagram of Figure \ref{fig:phase_diagram}. These results are shown in the phase diagram of Figure \ref{fig:diagFALQON}, in the corresponding region. This method, however, is not suitable for determining the transitions occurring for $\kappa > 0.5$. Other approaches, such as perturbative analyses \cite{PhysRevE.75.021105} and Quantum Machine Learning, Tensor Networks combined with Conformal Field Theory \cite{ceaANNI} have been used for addressing that region of the parameter space.

Nevertheless, by using the method described in Reference \cite{Sa_Barreto_2002}, we can apply FQA to determine the boundary of the antiphase in the phase diagram of Fig. \ref{fig:phase_diagram}. 
To identify the boundary of the antiphase, we conducted an analysis of points at which the first excited state changes its quantum number, indicating a phase configuration change. We initialized FQA with two different states: $\ket{\chi_1, 1, -, 1, 7}$ and $\ket{\chi_1, 1, 1, 0, -}$. The first state converges to the first excited state of $H_p$ in the paramagnetic phase in an 8-site chain, while the second state converges to the first excited state in the antiphase. Keeping $g$ values constant, we varied the parameter $\kappa$ to find the point at which the two states converged to the same expected energy value, indicating the crossing point and hence the phase transition. The transition line obtained by this method is presented in the corresponding region in the phase diagram of Fig. \ref{fig:diagFALQON}.

\subsection{Correlation Functions}

\begin{figure}
\centering
\includegraphics[width=1\linewidth]{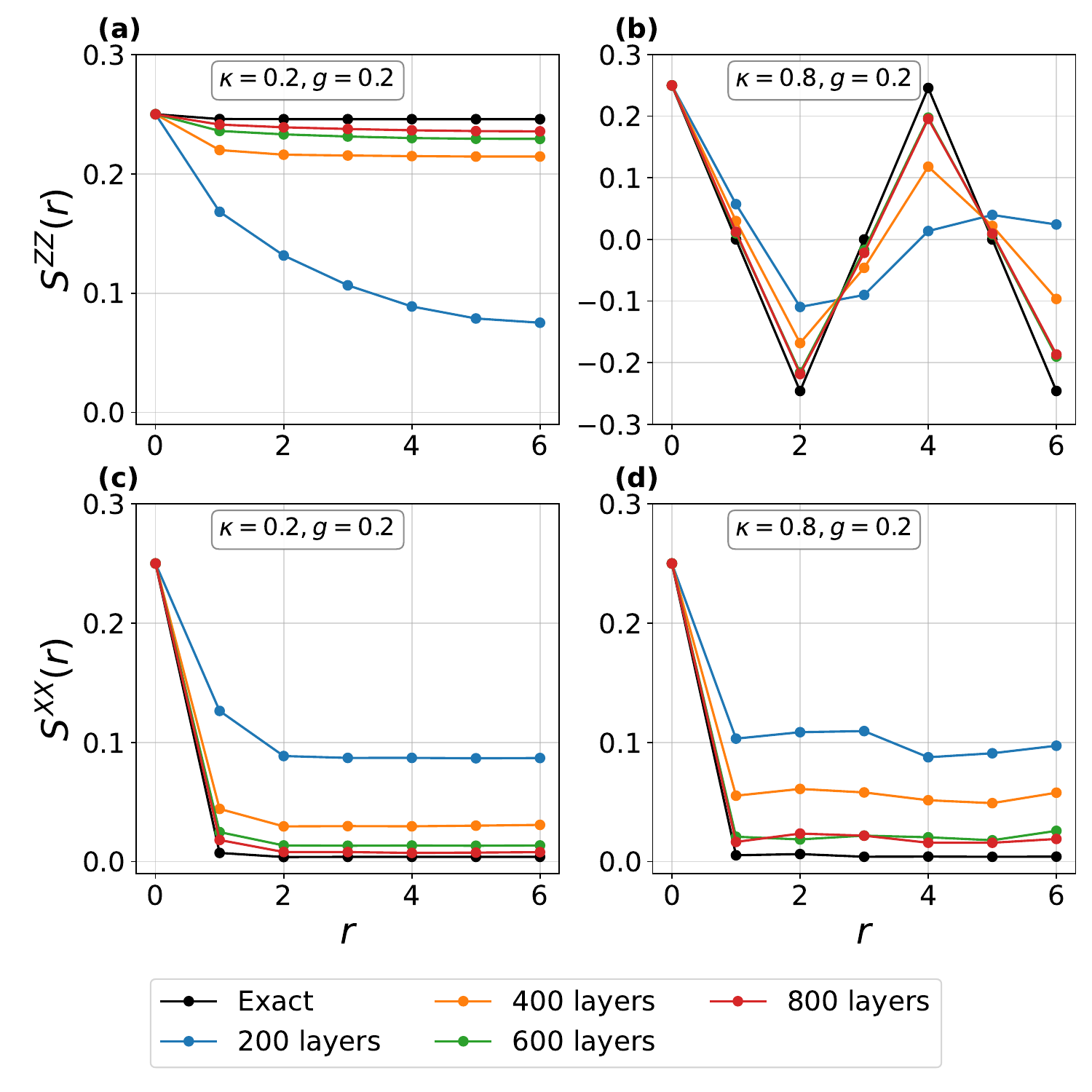}   
\caption{Real-space spin correlation functions of the ANNNI model for an 12-site chain, varying the number of layers of FQA, for different values of $\kappa$ and $g$. The simulations use $\Delta t = 0.02$. In (a) and (b), spatial dependence of the $zz$ correlation function, (a) $\kappa = 0.2, g = 0.2$, and (b) $\kappa = 0.8, g = 0.2$. In (c) and (d), spatial dependence of the $xx$ correlation function, (c) $\kappa = 0.2, g = 0.2$, and (d) $\kappa = 0.8, g = 0.2$.
}
\label{fig:corre}
\end{figure}

In this section, we used the ground state obtained with FQA to calculate the correlation functions, considering different values of $\kappa$ and $g$ that represent various phases of the model. Figure \ref{fig:corre} shows the spin correlation functions in real space according to Eq. (\ref{Smumu}), calculated using the ground state obtained through FQA for a 12-site chain ($L = 12$), varying the number of algorithm layers and comparing with exact diagonalization. In Figs. \ref{fig:corre}(a) and \ref{fig:corre}(b), we have the spin correlation function $S^{zz}(r)$ calculated for parameters (a) $\kappa = 0.2$, $g = 0.2$ and (b) $\kappa = 0.8$, $g = 0.2$. In Figs. \ref{fig:corre}(c) and \ref{fig:corre}(d), we have the spin correlation function $S^{xx}(r)$ calculated for parameters (c) $\kappa = 0.2$, $g = 0.2$ and (d) $\kappa = 0.8$, $g = 0.2$. It can be observed that as we increase the number of layers, the correlation function approaches the exact value.

Spin correlation functions provide important information about how spins are correlated at different positions along the chain. They describe the probability of finding two spins along the chain with specific orientations at a given separation $r$. To understand the magnetic properties, it is essential to analyze the behavior of correlations in different magnetic phases of the system. Distinct changes in these characteristics can indicate magnetic phase transitions, where the system transitions from one magnetic configuration to another. The analysis of spin correlation functions in real space allows not only to identify magnetic phases but also to understand the long-range properties and magnetic order present in the system.

\begin{figure}[h!]
\centering
\includegraphics[width=1\linewidth]{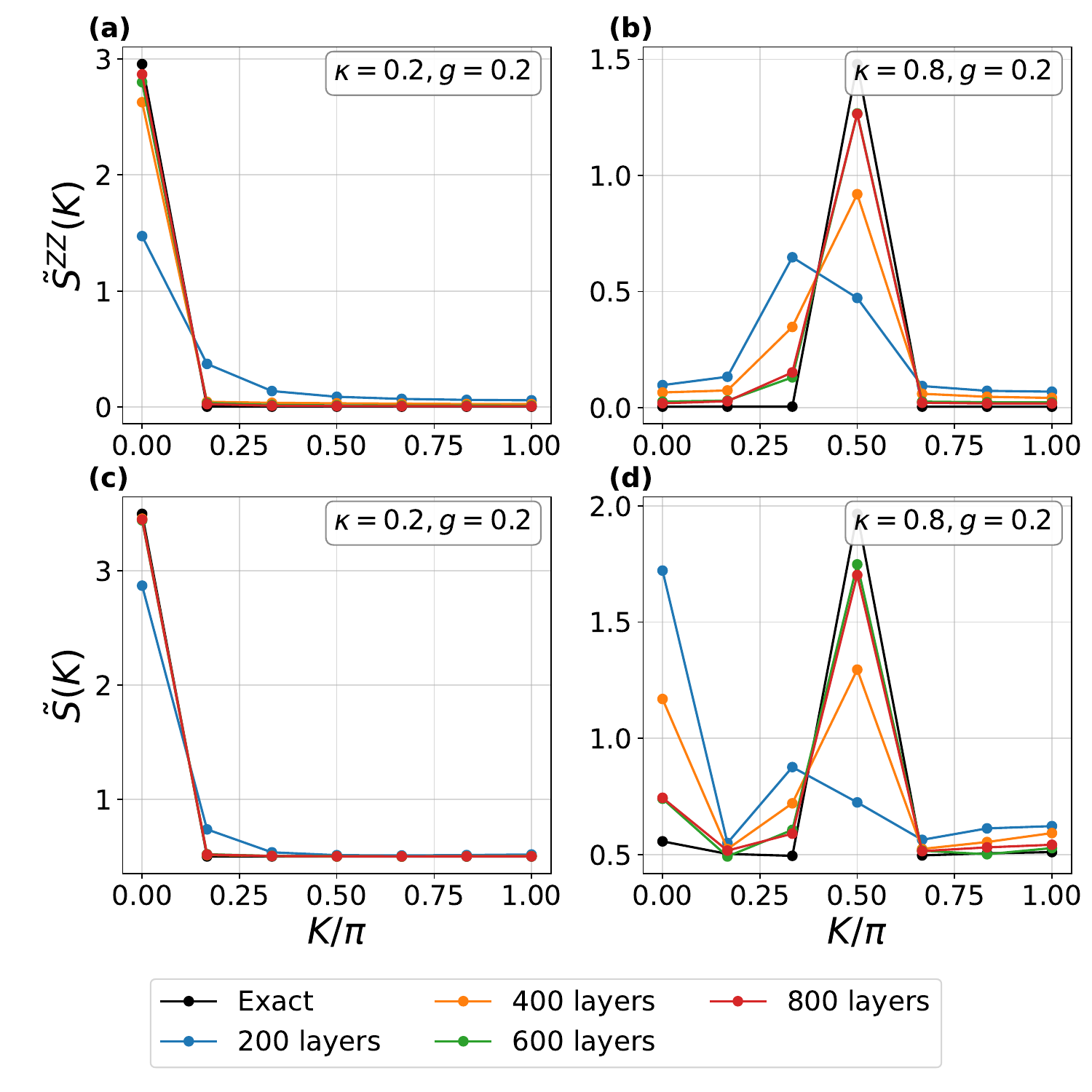} 
\caption{Magnetic structure factors of the ANNNI model for an 12-site chain, varying the number of layers of FQA, for different values of $\kappa$ and $g$. The simulations use $\Delta t = 0.02$. In (a) and (b), magnetic structure factor $\tilde{S}^{zz}(K)$, (a) $\kappa = 0.2$, $g = 0.2$, and (b) $\kappa = 0.8$, $g = 0.2$. In (c) and (d), the sum of the magnetic structure factors. In (c) $\kappa = 0.2$, $g = 0.2$, and in (d) $\kappa = 0.8$, $g = 0.2$.
}
\label{fig:FE}
\end{figure} 

\subsection{Magnetic Structure Factors}

Finally, we studied the magnetic structure factors and their sums. Using the ground state and Eqs. (\ref{Smuq}, \ref{sigmaq}), we calculated the structure factors and their sums for $K=0,...,L$ (see Appendix \ref{ap:FSS}). The magnetic structure factors in Figure \ref{fig:FE} provide a description of the spatial distribution of magnetic correlations along the one-dimensional chain, calculated for a 12-site chain with varying layers of the FQA algorithm. In Figs. \ref{fig:FE}(a) and \ref{fig:FE}(b), we have the factor $\tilde{S}^{zz}(K)$ calculated for the parameters (a) $\kappa = 0.2$; $g = 0.2$ and (b) $\kappa = 0.8$; $g = 0.2$. In Figs. \ref{fig:FE}(c) and \ref{fig:FE}(d), we have the sum of the factors $\tilde{S}(K)$ calculated for the parameters (c) $\kappa = 0.2$; $g = 0.2$ and (d) $\kappa = 0.8$; $g = 0.2$.

The magnetic structure factors $\tilde{S}^{\mu\mu}(K)$ reveal how spins at different positions along the chain are correlated. Different values of $K$ uncover various spatial scales associated with the system's magnetic properties. On the other hand, the sum of the magnetic structure factors provides a broader perspective on global magnetic properties, highlighting patterns common to all spin directions. This is essential for identifying magnetic order patterns in the model's ground state. By examining how $\tilde{S}(K)$ varies with $K$, specific magnetic characteristics of the system can be identified. For instance, peaks may indicate oscillation modes or well-defined magnetic orders, while valleys suggest regions with weaker magnetic correlations.

The study of magnetic structure factors and their sums is an important tool for exploring the magnetic properties of models such as the ANNNI. This analysis enables understanding both the local and global magnetic behavior of the system in its ground state, allowing the identification of magnetic phases and potential phase transitions present in the model.

\section{Conclusion}
\label{concl}

In this work, we investigated the potential of a feedback-based quantum algorithm to analyze the critical properties of the ANNNI model. Our approach enabled the calculation of both ground and excited states without relying on classical optimization methods, demonstrating the efficiency of the FQA algorithm. We analyzed FQA's behavior when initialized in states belonging to different symmetry sectors of the ANNNI model, including inversion, reflection, and translation symmetries. Our results show that, when FQA is initialized in a state from a specific sector, the algorithm converges to the lowest-energy eigenstate of the Hamiltonian within that subspace. In this way, FQA facilitates the exploration of higher-energy states by targeting sectors that do not contain the ground state, broadening its applicability in the study of excited states within the system. Furthermore, it provides a versatile tool for investigating other spin models beyond the ANNNI system, underscoring its potential as an effective approach for studying diverse quantum systems. Additionally, we demonstrated that even when the preparation of excited states is not required, understanding the system's symmetries is important to facilitate the preparation of the ground state. If the algorithm is initialized in a state outside the sector containing the ground state or in a superposition of states from different sectors, the FQA may fail to converge to the ground state.

With this approach, we conducted a detailed exploration that included the phase diagram, correlation functions, and structure factors of the ANNNI model. We employed the Finite Size Scaling method to study quantum phase transitions, analyzed spin correlations in the ground state to understand correlation functions, and used the Discrete Fourier Transform to examine magnetic structure factors. Our findings establish FQA as a tool to examine magnetic systems, offering valuable insights into the magnetic behavior and critical properties of the ANNNI model. Additionally, it suggests the applicability of FQA to study other quantum systems, highlighting the robustness and versatility of feedback-based quantum algorithms in exploring many-body quantum phenomena.

Also, it is important to emphasize that while FQA could potentially achieve promising results, it requires deeper quantum circuits that are more feasible on robust, fault-tolerant quantum computers. For effective execution in the NISQ era, advanced error correction and mitigation techniques will be essential to ensure satisfactory performance. Future research may extend this approach to other quantum systems, further validating the robustness and versatility of feedback-based quantum algorithms in studying many-body quantum phenomena.

\begin{acknowledgments}
F.F.F acknowledge support from Funda\c{c}{\~a}o de Amparo {\`a} Pesquisa do Estado de S{\~a}o Paulo (FAPESP), project number 2023/04987-6 and from ONR, project number N62909-24-1-2012. G. E. L. P. and L. A. M. R. acknowledge support from Coordenação de Aperfeiçoamento de Pessoal de Nível Superior (CAPES), Projects No. 88887.829788/2023-00 and No. 88887.829793/2023-00, respectively.
\end{acknowledgments}

\appendix

\section{Symmetries of the ANNNI Model}
\label{ap:simetrias}

The analysis of the one-dimensional ANNNI model with a transverse field is facilitated by investigating its symmetries. A symmetry is a transformation that, when applied to a physical system, does not alter its observable properties, making it invariant under this transformation.

Consider an operator $P$ representing a transformation acting on a chain of $L$ spin-$1/2$ sites. If this transformation is a symmetry of the system, then $P$ commutes with the system's Hamiltonian $H$ ($[P, H] = 0$). This implies that all subspaces of $P$ are invariant under the action of $H$ \cite{cohen1986quantum}, allowing the construction of a set of eigenvectors common to both $H$ and $P$. The eigenvectors of $H$ can be identified by the eigenvalues $\phi_p$ of $P$, providing an additional identification to the eigenstates of $H$ beyond their energy $E_h$. For simplicity, we refer to the eigenvalue $\phi_p$ simply as $p$, which will be the quantum number associated with $P$.

If the system has more than one symmetry $\{P_1, P_2, \dots, P_n\}$ and the operators representing them commute with each other, it is possible to construct a basis from the eigenvectors common to $H$ and these operators. In this basis, the representation of the Hamiltonian matrix takes a block-diagonal form \cite{malvezzi2003introduction, Jung_2020}, where each block corresponds to a vector subspace characterized by the quantum numbers $p_1, p_2, \dots, p_n$. This subspace is referred to as the sector of quantum numbers $p_1, p_2, \dots, p_n$. The vectors belonging to these subspaces are eigenvectors of the operators, represented by:

\begin{equation}
    \ket{h, p_1, p_2, \dots, p_n}.
\end{equation}

The invariance of the operators under the action of $H$ also extends to the temporal evolution of the system. If, at an initial time $t_0$, the system is in a common eigenstate of $P_1$, $P_2$, $\dots$, $P_n$, it will remain in that eigenstate, with the same eigenvalues, at a later time $t$ \cite{Sakurai_Napolitano_2020}. This is represented by the equation:

\begin{widetext}
\begin{equation}
\begin{aligned}
    P_1 \dots P_n \,U(t, t_0) \ket{h, p_1, p_2, \dots, p_n} & = U(t, t_0) \, P_1 \dots P_n \ket{h, p_1, p_2, \dots, p_n} \\
    & = \phi_1 \dots \phi_n \, U(t, t_0) \ket{h, p_1, p_2, \dots, p_n}
\end{aligned}
\end{equation}
\end{widetext}

\noindent
where $U(t, t_0)$ is the time evolution operator from $t_0$ to $t$. Thus, if at time $t_0$, the system is in a state belonging to a specific sector of quantum numbers, then after the temporal evolution up to time $t$, the state of the system will continue to belong to the same sector.

If $P$ is a cyclic symmetry of order $n$ ($\mathbb{Z}_n$) \cite{PhysRevD.103.125001}, meaning it has a period $n$, where applying $P$ $n$ times to a specific state results in the same initial state:

\begin{equation}
    P^{\, n} \ket{ \psi } = \ket{ \psi }
\end{equation}
where $\ket{\psi}$ is any state. Then the eigenvectors of $P$ can be expressed as:

\begin{equation}
    \ket{p} = C \sum_{j=0}^{r-1} \left(\phi_p^{-1} P \right)^j \ket{\psi}
\end{equation}
where $C$ is a normalization constant. The eigenvalues of $P$ are expressed as:

\begin{equation}
    \phi_p = e^{i \frac{2\pi}{L}p}, \; p = 0, 1, \dots, n-1
\end{equation}

If $\ket{p} = 0$, then it is not possible to form an eigenvector of $P$ with eigenvalue $\phi_p$ from the state $\ket{\psi}$. In the ANNNI model, three main symmetries are identified: spin inversion, spatial reflection, and translation. The spin inversion symmetry, represented by the operator $I$, acts on a state by flipping the spins at each site:

$$
\ket{\upa \upa \dna \upa \dna \dna \upa} \;\; \rightarrow \;\; \ket{\dna\dna \upa \dna \upa \upa \dna}.
$$

Being a cyclic symmetry of order two, it has eigenvalues $\phi_{p_I} = \pm 1$, with corresponding eigenvectors:

\begin{equation}
\label{eqn: autovectors_I}
\ket{j, p_I} = C[\mathbb{1} + \phi_{p_I} I] \ket{\psi}
\end{equation}
where $j$ is a degeneracy index.

The reflection symmetry is represented by the operator $R$ and flips the spin order about a central reflection point in the chain:

$$
R \vert d_0 \, d_1 \, \cdots \, d_{L-2} \, d_{L-1} \rangle = \vert d_{L-1} \, d_{L-2} \, \cdots \, d_1 \, d_0 \rangle.
$$

Similar to $I$, $R$ is a cyclic symmetry of order two. Its eigenvalues are $\phi_{p_R} = \pm 1$, with corresponding eigenvectors:

\begin{equation}
\label{eqn: autovectors_R}
\ket{j, p_R} = C[\mathbb{1} + \phi_{p_R} R] \vert d_0 \, d_1 \, d_2 \, \cdots \, d_{L-1} \rangle.
\end{equation}

The translation symmetry is represented by the operator $T$. When acting on the chain, it shifts the site positions cyclically:
\begin{equation}
T \vert d_0 \, d_1 \, \cdots \, d_{L-2} \, d_{L-1} \rangle = \vert d_{1} \, d_{2} \, \cdots \, d_{L-1} \, d_0 \rangle.
\end{equation}

$T$ is a cyclic symmetry with an order equal to the chain size. Its eigenvalues are $\phi_{p_T} = e^{i\frac{2\pi}{L}{p_T}}$, where $p_T=0,1,\cdots,L-1$. The corresponding eigenvectors are:

\begin{equation}
\label{eqn:autovectors_T}
\ket{j, p_T} = C[\mathbb{1} + \phi^{-1}_{p_T} T + \cdots + (\phi^{-1}_{p_T} T)^{L-1}] \vert \psi \rangle.
\end{equation}

Although the operators $I, R$, and $T$ commute with the Hamiltonian of the model, it is not possible to form a set of operators that commute with each other, as the reflection operator does not commute with the translation operator for all chain sizes. To overcome this issue, we can use one of the powers of $T$ in place of $T$ directly. In chains with an even number of sites, we can use the operator $T^{L/2}$ as a substitute, which has eigenvalues $\phi_{p_{T^{L/2}}} = \pm 1$ with eigenvectors:

\begin{equation}
    \label{eq:autovectors_T2}
    \ket{j, p_{T^{L/2}}} = C[\mathbb{1} + \phi_{p_{T^{L/2}}}^{-1} T^{L/2}] \ket{\psi}.
\end{equation}

This way, the eigenvectors of the Hamiltonian are identified by the quantum numbers $p_I, p_R, p_{T^{L/2}}$. In certain system sectors, the translation operator commutes with the reflection operator, allowing these sectors to be subdivided into smaller sectors further characterized by $p_T$. Thus, the eigenstates within a given sector can be represented as:
\begin{equation}
    \ket{h, p_I, p_R, p_{T^2}, p_T}
\end{equation}
where $h$ acts as an identifier to distinguish eigenstates within the same sector. For states lacking specific quantum numbers, the symbol $(-)$ is used in the respective position to indicate the absence of that quantum number.

As an example, let's construct the basis of common eigenstates to the operators $I$, $R$, and $T^{L/2}$, considering a chain of 4 sites. To do this, we will successively apply Eqs. (\ref{eqn: autovectors_I}, \ref{eqn: autovectors_R}, \ref{eq:autovectors_T2}) to the states of the standard basis $\{\ket{\upa}, \ket{\dna}\}^{\otimes4}$. For the eigenstates that form sectors where the translation operator commutes with the reflection operator, we will additionally apply Eq. (\ref{eqn:autovectors_T}). This gives the following states:

\begin{widetext}
\begin{equation}
\begin{aligned}
    \ket{1, 0, 0, 0, 0} &=\dfrac{1}{\sqrt{2}} (\ket{\upa\upa\upa\upa} + \ket{\dna\dna\dna\dna}) \\
    \ket{2, 0, 0, 0, 0} &=\dfrac{1}{\sqrt{8}} (\ket{\dna\dna\upa\dna} + \ket{\upa\upa\dna\upa} + \ket{\upa\dna\dna\dna} + \ket{\dna\upa\upa\upa}  + \ket{\dna\upa\dna\dna} + \ket{\upa\dna\upa\upa} + \ket{\dna\dna\dna\upa} + \ket{\upa\upa\upa\dna}) \\
    \ket{3, 0, 0, 0, 0} &= \dfrac{1}{2} (\ket{\upa\upa\dna\dna} + \ket{\dna\dna\upa\upa} + \ket{\upa\dna\dna\upa} + \ket{\dna\upa\upa\dna}) \\
    \ket{4, 0, 0, 0, 0} &= \dfrac{1}{\sqrt{2}} (\ket{\upa\dna\upa\dna} + \ket{\dna\upa\dna\upa}) \\
    \ket{1, 0, 0, 0, 2} &= \dfrac{1}{2} (\ket{\upa\upa\dna\dna} + \ket{\dna\dna\upa\upa} - \ket{\upa\dna\dna\upa} - \ket{\dna\upa\upa\dna}) \\
    \ket{1, 1, 0, 0, 0} &= \dfrac{1}{\sqrt{2}} (\ket{\upa\upa\upa\upa} - \ket{\dna\dna\dna\dna})\\
    \ket{2, 1, 0, 0, 0} &= \dfrac{1}{\sqrt{8}} (\ket{\upa\upa\upa\dna} + \ket{\upa\upa\dna\upa} + \ket{\upa\dna\upa\upa} - \ket{\upa\dna\dna\dna} + \ket{\dna\upa\upa\upa} - \ket{\dna\upa\dna\dna} - \ket{\dna\dna\upa\dna} - \ket{\dna\dna\dna\upa})\\
    \ket{1, 0, 1, 0, 2} &= \dfrac{1}{\sqrt{8}} (\ket{\upa\upa\upa\dna} - \ket{\upa\upa\dna\upa} + \ket{\upa\dna\upa\upa} - \ket{\upa\dna\dna\dna} - \ket{\dna\upa\upa\upa} + \ket{\dna\upa\dna\dna} - \ket{\dna\dna\upa\dna} + \ket{\dna\dna\dna\upa})\\
    \ket{1, 1, 1, 0, 2} &= \dfrac{1}{\sqrt{8}} (\ket{\upa\upa\upa\dna} - \ket{\upa\upa\dna\upa} + \ket{\upa\dna\upa\upa} + \ket{\upa\dna\dna\dna} - \ket{\dna\upa\upa\upa} - \ket{\dna\upa\dna\dna} + \ket{\dna\dna\upa\dna} - \ket{\dna\dna\dna\upa}) \\
    \ket{2, 1, 1, 0, 2} &= \dfrac{1}{\sqrt{2}} (\ket{\upa\dna\upa\dna} - \ket{\dna\upa\dna\upa}) \\
    \ket{1, 0, 0, 1, -} &= \dfrac{1}{\sqrt{8}} (\ket{\upa\upa\upa\dna} - \ket{\upa\upa\dna\upa} - \ket{\upa\dna\upa\upa} + \ket{\upa\dna\dna\dna} + \ket{\dna\upa\upa\upa} - \ket{\dna\upa\dna\dna} - \ket{\dna\dna\upa\dna} + \ket{\dna\dna\dna\upa}) \\
    \ket{1, 1, 0, 1, -} &= \dfrac{1}{\sqrt{8}} (\ket{\upa\upa\upa\dna} - \ket{\upa\upa\dna\upa} - \ket{\upa\dna\upa\upa} - \ket{\upa\dna\dna\dna} + \ket{\dna\upa\upa\upa} + \ket{\dna\upa\dna\dna} + \ket{\dna\dna\upa\dna} - \ket{\dna\dna\dna\upa}) \\
    \ket{2, 1, 0, 1, -} &= \dfrac{1}{\sqrt{2}} (\ket{\upa\dna\dna\upa} - \ket{\dna\upa\upa\dna}) \\
    \ket{1, 0, 1, 1, -} &= \dfrac{1}{\sqrt{8}} (\ket{\upa\upa\upa\dna} + \ket{\upa\upa\dna\upa} - \ket{\upa\dna\upa\upa} - \ket{\upa\dna\dna\dna} - \ket{\dna\upa\upa\upa} - \ket{\dna\upa\dna\dna} + \ket{\dna\dna\upa\dna} + \ket{\dna\dna\dna\upa}) \\
    \ket{1, 1, 1, 1, -} &= \dfrac{1}{\sqrt{8}} (\ket{\upa\upa\upa\dna} + \ket{\upa\upa\dna\upa} - \ket{\upa\dna\upa\upa} + \ket{\upa\dna\dna\dna} - \ket{\dna\upa\upa\upa} + \ket{\dna\upa\dna\dna} - \ket{\dna\dna\upa\dna} - \ket{\dna\dna\dna\upa}) \\
    \ket{2, 1, 1, 1, -} &= \dfrac{1}{\sqrt{2}} (\ket{\upa\upa\dna\dna} - \ket{\dna\dna\upa\upa}) \\
\end{aligned}
\label{eq:base_bloco_diagonal}
\end{equation}
\end{widetext}

\begin{figure}[!h]
    \centering
    \includegraphics[width=1\linewidth]{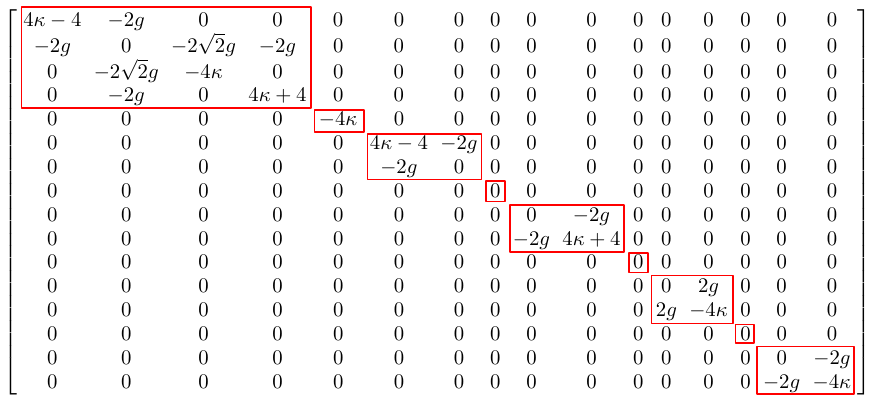}
    \caption{{Block-diagonalization of the Hamiltonian of the ANNNI model for generic parameters $\kappa$ and $g$, obtained using the basis defined in (\ref{eq:base_bloco_diagonal}).}}
    \label{fig:Hp_bloco_diagonal}
\end{figure}

By changing the Hamiltonian basis to the one defined in (\ref{eq:base_bloco_diagonal}), we obtain a block-diagonal representation, as shown in Figure \ref{fig:Hp_bloco_diagonal}. Table \ref{tab:espectro} presents the spectrum of the Hamiltonian of the ANNNI model for a 4-site chain, obtained through exact diagonalization of each block in the representation using the specified basis. Two distinct parameter sets are considered: $\kappa = 0.2, g = 0.2$ and $\kappa = 0.8, g = 0.2$. Each row in the table corresponds to a specific configuration of quantum numbers characterizing the sector, along with the eigenvalues associated with these configurations for the two parameter sets.

\begin{table}[!h]
\centering
\caption{Spectrum of the Hamiltonian of the ANNNI model for a 4-site chain, with different parameter sets ($\kappa$ and $g$). The quantum numbers $p_I$, $p_R$, $p_{T^2}$, and $p_T$ indicate inversion, reflection, translation squared, and translation transformations, respectively. The energies of the eigenstates associated with these configurations are shown for the two parameter sets. When a quantum number is not present, it is represented by "-", indicating the absence of that transformation for the corresponding state.}
\resizebox{\columnwidth}{!}{%
\begin{tabular}{ccccccc}
\hline\hline
$p_I$ & $p_R$ & $p_{T^2}$ & $p_T$ & $E(\kappa = 0.2, g = 0.2)$ & $E(\kappa = 0.8, g = 0.2)$ \\ \hline
0     & 0     & 0         & 0     & -3.2516                    & -3.2994                    \\ 
0     & 0     & 0         & 0     & -1.0829                    & -0.9493                    \\ 
0     & 0     & 0         & 0     & 0.3008                     & 0.2263                     \\ 
0     & 0     & 0         & 0     & 4.8336                     & 7.2223                     \\ 
0     & 0     & 0         & 2     & -0.8000                    & -3.2000                    \\ 
1     & 0     & 0         & 0     & -3.2492                    & -0.9657                    \\ 
1     & 0     & 0         & 0     & 0.0492                     & 0.1657                     \\ 
0     & 1     & 0         & 2     & 0.0000                     & 0.0000                     \\ 
1     & 1     & 0         & 2     & -0.0331                    & -0.0222                    \\ 
1     & 1     & 0         & 2     & 4.8331                     & 7.2222                     \\ 
0     & 0     & 1         & -     & 0.0000                     & 0.0000                     \\ 
1     & 0     & 1         & -     & -0.9657                    & -3.2492                    \\ 
1     & 0     & 1         & -     & 0.1657                     & 0.0492                     \\ 
0     & 1     & 1         & -     & 0.0000                     & 0.0000                     \\ 
1     & 1     & 1         & -     & -0.9657                    & -3.2492                    \\ 
1     & 1     & 1         & -     & 0.1657                     & 0.0492                     \\ \hline\hline
\end{tabular}%
}
\label{tab:espectro}
\end{table}

\section{Convergence of FQA in Different Symmetry Sectors}\label{ap:convergence}

In this appendix, we analyze how the selection of distinct symmetry sectors for the preparation of initial states affects the convergence of FQA. Additionally, we examine the convergence for different states within the same sector.

\begin{figure}[!h]
\centering
\includegraphics[width=1\linewidth]{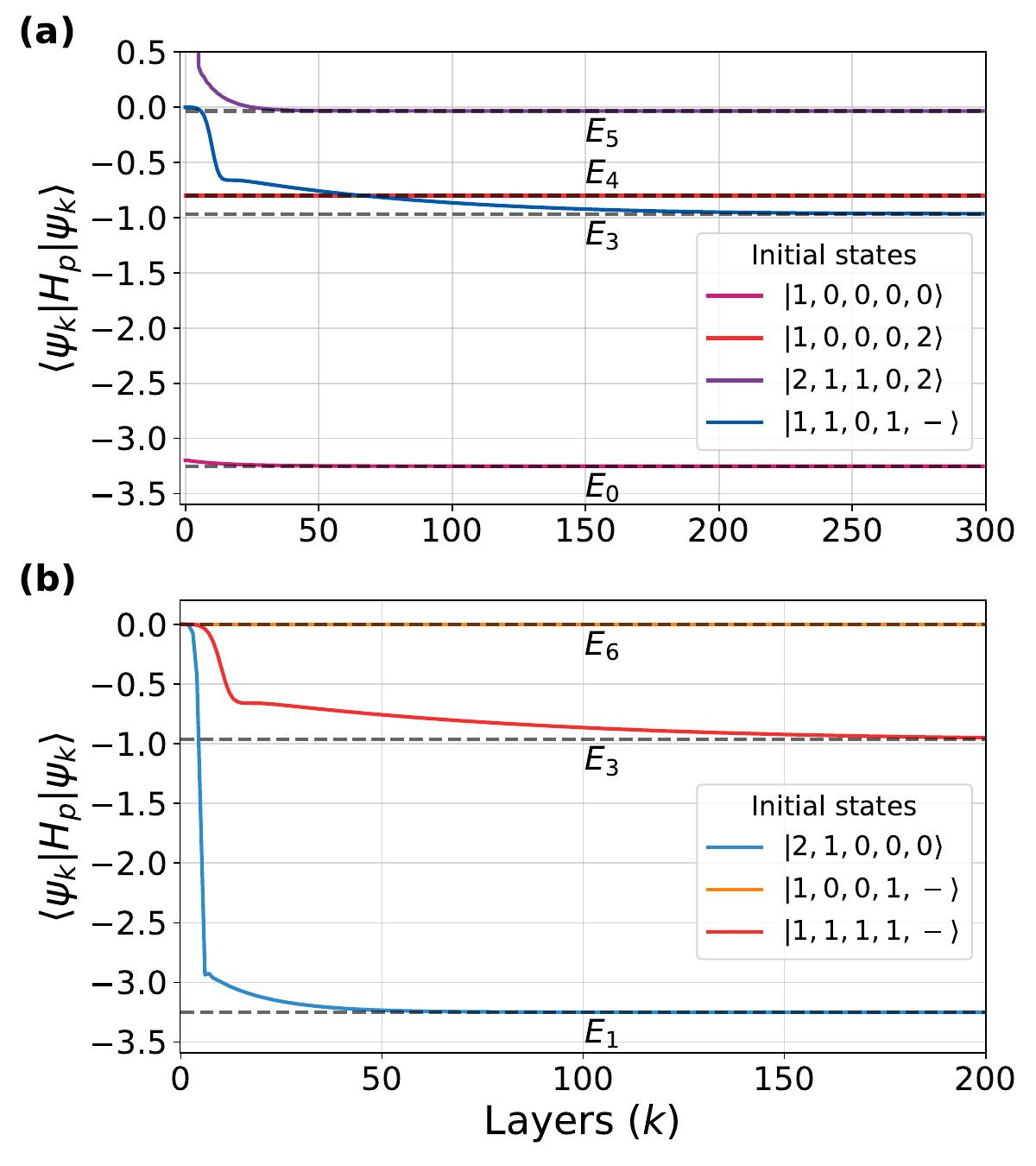}  
\caption{{
Numerical simulations of FQA applied to the ANNNI model for a 4-site chain, with parameters $\kappa = 0.2$ and $g = 0.2$ and time step $\Delta t = 0.06$. In graph (a), the evolution of initial states toward energies $E_0$, $E_3$, $E_4$, and $E_5$ is observed, while in graph (b), convergence occurs for $E_1$, $E_3$, and $E_6$. Dashed lines indicate the energy levels of the ANNNI Hamiltonian, and the curves represent the convergence of the cost function $J = \langle\Psi_{k}|H_p|\Psi_{k}\rangle$ as a function of layer $k$, with qubits initialized in different eigenvalues as specified in Eq. (\ref{eq:base_bloco_diagonal}).}}
\label{fig:graf_convergencia_dif_simetrias}
\end{figure}

Figure \ref{fig:graf_convergencia_dif_simetrias} shows the convergence of FQA applied to the ANNNI model, using this Hamiltonian as $H_p$ in a 4-site chain, with parameters $\kappa = 0.2$ and $g = 0.2$. The simulation considers initial states belonging to different symmetry sectors, listed in (\ref{eq:base_bloco_diagonal}). Observing the graph and the energies in Table \ref{tab:espectro}, it is noted that the objective function tends to converge to the lowest energy in the sector where the initial state is located. For sectors containing only one eigenstate, the objective function remains constant at that energy value, as the initial state is already the eigenstate of $H_p$ for that sector.

\begin{figure}[!h]
\centering
\includegraphics[width=1\linewidth]{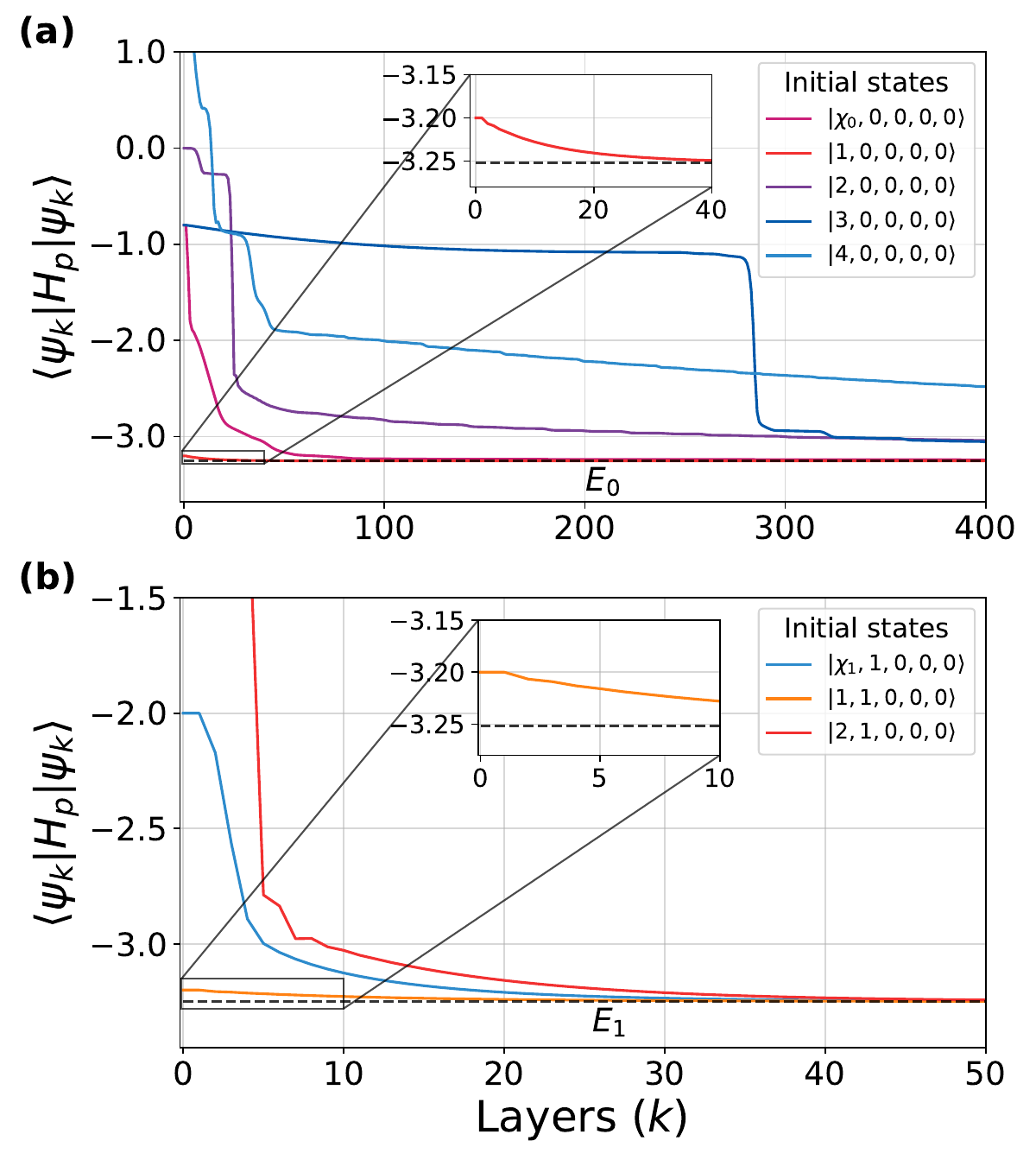}  
\caption{{
Numerical simulations of FQA applied to the ANNNI model for a 4-site chain, with parameters $\kappa = 0.2$ and $g = 0.2$, and time step $\Delta t = 0.06$. In (a), the algorithm converges to the ground state, while in (b), it converges to the first excited state. Dashed lines indicate the energy levels of the ANNNI Hamiltonian, and the curves show the convergence of the cost function $J = \langle\Psi_{k}|H_p|\Psi_{k}\rangle$ as a function of layer $k$, with qubits initialized in different eigenvalues as specified in Eq. (\ref{eq:base_bloco_diagonal}).}}
\label{fig:graf_convergencia_estados}
\end{figure}

Figure \ref{fig:graf_convergencia_estados} presents the convergence of different states within the same sector toward the lowest corresponding eigenstate. Figure \ref{fig:graf_convergencia_estados}(a) illustrates convergence in the sector containing the ground state of $H_p$, while Fig. \ref{fig:graf_convergencia_estados}(b) shows convergence in the first excited state sector. For the simulations in Fig. \ref{fig:graf_convergencia_estados}, the same parameters ($\kappa = 0.2$ and $g = 0.2$) were used, positioning the system in the ferromagnetic regime, where the gap between the ground state and the first excited state is small. It is observed that different states follow distinct trajectories in Hilbert space, resulting in varied convergence patterns and altering the number of layers required to reach the desired state. In Fig. \ref{fig:graf_convergencia_estados}(a), the state that converged most quickly was $\ket{1, 0, 0, 0, 0} = (\ket{0000} + \ket{1111})/\sqrt{2}$, while in Fig. \ref{fig:graf_convergencia_estados}(b), the most efficient state was $\ket{1, 1, 0, 0, 0} = (\ket{0000} - \ket{1111})/\sqrt{2}$.

\section{Finite Size Scaling and Magnetic Properties}
\label{ap:FSS}

The Finite Size Scaling (FSS) method, introduced by Fisher and Barber \cite{Fisher, Barber}, is a powerful tool in statistical physics and the theory of critical systems. It is often used to study the behavior of physical systems around critical points in finite dimensions, especially in systems with phase transitions. The fundamental idea behind FSS is that the physical properties of a system depend on its size in a predictable and universal manner.

The energy gap of the quantum Hamiltonian $\Delta (\kappa,g)$ as a function of parameters $(\kappa,g)$ is inversely proportional to the correlation length $\xi$. Thus, in the thermodynamic limit, as the chain size $L \rightarrow \infty$, the energy gap tends to zero at the critical point, $\Delta (\kappa,g) \rightarrow 0$, leading to a singularity where the correlation length diverges, $\xi \rightarrow \infty$. At the transition point, we have $\xi \approx L$, so that we obtain

\begin{equation}
    \frac{L}{\xi } = L\cdot \Delta(\kappa,g) = L\cdot (E_{1}(\kappa,g)- E_{0}(\kappa,g)) \rightarrow 1,
    \label{eq:FSS}
\end{equation}

\noindent
where $E_{0}(\kappa,g)$ is the ground state energy and $E_{1}(\kappa,g)$ is the energy of the first excited state calculated at point ($\kappa,g$) for a chain of $L$ sites. An estimate for the critical point (phase transition point) can be obtained when the function $L\cdot (E_{1}(\kappa,g)-E_{0}(\kappa,g))$ coincides for two adjacent chain sizes, $L$ and $L+1$. In practical terms, we plot the function $L\cdot (E_{1}(\kappa,g)-E_{0}(\kappa,g))$ while fixing one parameter and varying the other. When the plots for two adjacent chain sizes overlap, the intersection point of these curves provides the estimate of the critical point. By repeating this process for successive pairs of larger chains, we obtain a sequence of estimates that converge to the value of the critical point as $L \rightarrow \infty$.

To investigate the magnetic properties obtainable exclusively from the ground state in the context of the ANNNI model, our aim was to explore the intricate relationships of the correlation functions that emerge, highlighting the nuances and implications these correlations have on the magnetic properties of the system.

Understanding the ground state allows us to compute the magnetic properties associated with each of the magnetic phases of the system, characterized by spin correlation functions in real space,
\begin{equation}
S^{\mu\mu}(r)=\langle  S^{\mu}_i S^{\mu}_{i+r} \rangle,\ \mu=x,y,z\;,
\label{Smumu}
\end{equation}
and their corresponding magnetic structure factors via the Discrete Fourier Transform,
\begin{equation}
\tilde{S}^{\mu\mu}(K)= \frac{1}{L}\sum_{i,j=1}^L{\rm e}^{\imath K(i-j)} \langle  S^{\mu}_i S^{\mu}_{j} \rangle,\ \mu=x,y,z\;,
\label{Smuq}
\end{equation}
as well as their sums,
\begin{equation}
\tilde{S}(K)= \sum_{\mu=x,y,z} \tilde{S}^{\mu\mu}(K)\;,
\label{sigmaq}
\end{equation}
where $\langle  \ldots \rangle\equiv \langle \psi_0|  \ldots|\psi_0\rangle$ denotes the average value in the ground state, $S^{\mu}_i = \frac{1}{2}\sigma_i^{\mu}$ is the spin operator component at site $i$ (assuming $\hbar = 1$), $r$ is the distance between the sites where the correlation is being measured, and $K$ (in units of $2\pi/L$) is the wave number characterizing the periodicity of magnetic ordering.

\section{Quantum Lyapunov Control}

The quantum Lyapunov control framework is utilized to design one or more time-dependent controls aiming to asymptotically guide a quantum dynamical system towards a desired state \cite{Hou_2012}. To introduce this framework, we start with a quantum system governed by the time-dependent Schrödinger equation (assuming $\hbar$ = 1),
\begin{equation}
i\frac{d}{dt}|\Psi(t)\rangle=[H_p+H_d\beta(t)]|\Psi(t)\rangle, \label{se1}
\end{equation}
where $H_p$ represents the time-independent part of the Hamiltonian, referred to as the problem Hamiltonian. On the other hand, $H_d$ denotes the control Hamiltonian, which couples a time-dependent control function $\beta(t)$ to the system; this is termed the driver Hamiltonian. The extension to cases with multiple control functions is straightforward, as discussed, for example, in Ref. \cite{2Magann_2022}.

The quantum control problem addresses the projection of $\beta(t)$ to guide $|\Psi(t)\rangle$ towards a state that minimizes an objective function $J$. Quantum Lyapunov control has been developed for various distinct objective functions, such as capturing the distance to a target state \cite{31,32,33}, the error relative to the target state \cite{34}, and the expected value of a target observable. In this context, we focus on the observable case and define $J$ as the expected value of $H_p$,
\begin{equation}
    J = \langle\Psi(t)|H_{p}|\Psi(t)\rangle. \label{se2}
\end{equation}

To solve the control problem, the Lyapunov control method seeks a function $\beta(t)$ that leads to a monotonically decreasing objective function $J$ over time $t$. This method is asymptotic, eliminating the need to specify an endpoint in advance. Particularly, we seek a $\beta(t)$ function that satisfies the derivative condition:
\begin{equation}
\frac{dJ}{dt} \leq 0, \forall t. \label{se3}
\end{equation}
Evaluating the left-hand side of Eq. (\ref{se3}) using Eqs. (\ref{se1}) and (\ref{se2}), we obtain:
\begin{equation}
\frac{dJ}{dt} = A(t)\beta (t), \label{se4}
\end{equation}
where $A(t)$ abbreviates the time-dependent expectation value $A(t)\equiv \langle\Psi(t)|i[H_{d},H_{p}]|\Psi(t)\rangle$. The derivative condition in Eq. (\ref{se3}) can be satisfied by choosing $\beta (t) = wf(t,A(t))$, where $w > 0$ is a positive weight and $f(t,A(t))$ is chosen such that $f(t,0) = 0$ and $A(t)f(t,A(t)) > 0$ for all $A(t) \neq 0$. The specific formulation used in this article considers $w = 1$ and $f(t,A(t)) = A(t)$, resulting in the following control law:
\begin{equation}
\beta (t) = -A(t). \label{se5}
\end{equation}
This control law satisfies the condition set by Eq. (\ref{se3}), resulting in $\frac{dJ}{dt} = -(A(t))^{2} \leq 0$ for all $A(t)$.

The convergence of quantum Lyapunov control has been analyzed using the LaSalle's invariance principle \cite{36}, which identified a set of sufficient conditions to ensure asymptotic convergence to the global minimum of Eq. (\ref{se2}). However, these conditions are often stringent and rarely met in practical applications. Nonetheless, convergence to the global minimum is frequently observed in practice, as demonstrated by numerical simulations \cite{1Magann_2022, larsen2023feedbackbased}, even in scenarios where theoretical convergence criteria are not satisfied. Better understanding the necessary conditions for the convergence of quantum Lyapunov control, in order to bridge the gap between mathematical results and numerical observations, remains an open and intriguing research challenge.

\bibliography{main}

\end{document}